\documentclass[a4paper,fleqn,usenatbib]{mnras}

\usepackage{newtxtext,newtxmath}

\usepackage[T1]{fontenc}
\usepackage{ae,aecompl}

\usepackage{graphicx}	
\usepackage{amsmath}	
\usepackage{amssymb}	

\usepackage{tabularx}

\title[Evolution of WR stars and SN Ib/Ic progenitors]{Towards a better understanding of the evolution of Wolf-Rayet
stars and Type Ib/Ic supernova progenitors}
\author[S.-C. Yoon]{
Sung-Chul Yoon$^{1,2}$\thanks{yoon@astro.snu.ac.kr}
\\
$^{1}$Department of Physics and Astronomy, Seoul National University, Gwanak-ro 1, Gwanak-gu, Seoul, 08826, South Korea \\
$^{2}$Monash Centre for Astrophysics, School of Physics and Astronomy, Monash University, Victoria 3800, Australia
}

\date{Accepted XXX. Received YYY; in original form ZZZ}

\pubyear{2017}

\begin{document}
\label{firstpage}
\pagerange{\pageref{firstpage}--\pageref{lastpage}}
\maketitle

\begin{abstract}
Hydrogen-deficient Wolf-Rayet (WR) stars are potential candidates of Type Ib/Ic
supernova (SN Ib/Ic)  progenitors and their evolution is governed by mass loss.
Stellar evolution models with the most popular prescription for  WR mass-loss rates given by Nugis \& Lamers
have difficulties in explaining the luminosity distribution of WR stars of WC and
WO types and the SN Ic progenitor properties. Here we suggest some improvements
in the WR mass-loss rate prescription and discuss its implications for the evolution 
of WR stars and SN Ib/Ic progenitors.    
Recent studies on
Galactic WR stars clearly indicate that the mass-loss
rates of WC stars are systematically higher than those of WNE stars for a given
luminosity. The luminosity and initial metallicity dependencies of WNE mass-loss rates are also significantly
different from  those of WC stars.  These
factors have not been adequately considered together in previous stellar evolution models.
We also find that an overall increase of WR mass loss rates by about 60 per cent
compared to the empirical values obtained with a clumping
factor of 10 is needed to explain the most faint WC/WO stars. This moderate
increase with our new WR mas-loss rate prescription results in SN Ib/Ic progenitor models
more consistent with observations than those given by the Nugis \& Lamers prescription. 
In particular, our new models predict that the properties of SN Ib and SN Ic progenitors
are distinctively different, rather than they form a continuous sequence. 
\end{abstract}


\section{Introduction}\label{ref:introduction}

Massive stars may lose their hydrogen-rich envelopes during the post-main
sequence phase via stellar winds~\citep{Conti76}.  Thus-formed
hydrogen-deficient stars are usually observed as classical Wolf-Rayet (WR)
stars having high bolometric luminosities of $\log L \gtrsim 5.0$ and strong
emission lines resulting from optically thick winds \citep[see][for a review]{Crowther07}. If they
can produce  supernovae at their death, they would appear as Type Ib or Ic
supernovae (SNe Ib/Ic).  Hydrogen-deficient stars as SN Ib/Ic progenitors may
also be produced in interacting binary systems, and they can be 
less massive than classical WR stars that typically have $M >
10~{M_\odot}$~\citep{Podsiadlowski92, Vanbeveren98, Wellstein99, Yoon10,
Eldridge13, Yoon17}.  The quasi-WR star HD45166 is one such
candidate~\citep[e.g.,][]{Steiner05}.  

WR stars lose mass via radiation-driven winds at very high rates ($\dot{M}
\gtrsim 10^{-5}~{M_\odot~\mathrm{yr^{-1}}}$) and their evolution towards core collapse is
critically determined by mass loss.  In spite of both observational and
theoretical efforts during the last three decades, the WR mass-loss rates at
various stages are still subject to considerable uncertainty.  The
prescriptions of WR mass-loss rates used in stellar evolution models before
2000 gave very large values~\citep[e.g.,][]{Maeder87, Langer89, Hamann95}. One
of the most striking results in these studies was the prediction that even very
massive stars with $M_\mathrm{ZAMS} \approx 60~{M_\odot}$  may have a final
mass as low as 3.0 -- 4.0~${M_\odot}$, whether or not they are in binary
systems \citep{Woosley93, Wellstein99}. 

However, consideration of  wind clumping
~\citep[e.g.,][]{Moffat94, Lepine99} in later empirical estimates of WR
mass-loss rates resulted in much lower values~\citep[e.g.,][]{Hamann98,
Hamann06, Crowther07, Sander12}. For example, currently the most popular
prescription by \citet[][hereafter, NL]{Nugis00} gives WR mass-loss rates
almost 10 times lower than those used in the above-quoted theoretical studies.  Stellar
evolution models adopting the NL prescription at solar metallicity predict much higher
final masses of WR stars ($M_\mathrm{f} > 10~{M_\odot}$; e.g.,
\citealt{Meynet05, Eldridge06, Georgy12}) than predicted by the models of the 80s
and 90s. This implies that single massive stars at $Z \lesssim Z_\odot$
would not produce ordinary SNe Ib/Ic, of which the ejecta
masses have been inferred to be lower than about $6~{M_\odot}$ for most
cases~\citep{Drout11, Cano13, Taddia15, Lyman16}. Binary star evolution models
at solar metallicity using the NL prescription also have great difficulties in
explaining SN Ic progenitors in terms of helium and ejecta masses, although they can
explain the overall properties of SN Ib progenitors that are
helium-rich~~\citep[see][for a recent review]{Yoon15}. 

Recent studies by the Potsdam group  presents a homogeneous set of WR star
properties for both WN and WC types, which gives an excellent observational
constraint on WR star properties~\citep{Hamann06, Sander12, Hainich14}.  As
discussed below, the observed population of WN/WC stars cannot be well
explained by the models using the NL prescription either. In particular, many
WC stars appear to be too faint compared to the prediction of stellar evolution
models~\citep{Sander12}. 

Given that stellar evolution models with the NL prescription can properly
predict neither  WR star population nor SN Ic progenitors, we need to consider
revising the prescription for WR mass-loss rates.  Recently,
\citet[][hereafeter, TSK]{Tramper16} presented a new prescription for mass-loss rates of
hydrogen-free WR stars. They argued that the dependencies of empirical WR mass-loss rates
on the luminosity and surface helium abundance are weaker than those of the NL
prescription, for WC and WO stars.  In this paper, we further discuss this
issue on the dependencies of WR mass-loss rates on physical parameters.  We argue
that the combination of the empirical mass-loss rates of WNE stars and the TSK
prescription provides a better qualitative agreement with the observed mass
loss rates of hydrogen-free WR stars than the TSK prescription alone and 
the standard NL prescription.  We also argue for the need of an overall
increase of WR mass-loss rates by about 60 per cent compared to the values commonly
used in recent stellar evolution models. 

In Section~\ref{sect:wind}, we compare the mass-loss rate prescriptions of NL and
TSK, and confront them with the Potsdam WR sample. We suggest a new prescription
for WR mass-loss rates that may better reflect the qualitative features of the
empirical mass-loss rates of hydrogen-free WR stars.  In
Section~\ref{sect:evol}, we present new evolutionary models of pure helium
stars (He stars) using this prescription and compare them with
observations. We also show that an overall increase of WR mass-loss rates
is needed to explain the luminosity distribution of WC/WO stars.
We discuss its implications for SN Ib/Ic progenitors in
Section~\ref{sect:snIbc}, and the issue of temperature discrepancy between
models and observations of WR stars in Section~\ref{sect:temperature}.  We
conclude our paper in Section~\ref{sect:conclusion}.

\section{Mass loss rates of WR stars}\label{sect:wind}

\begin{figure}
\begin{center}
\includegraphics[width = \columnwidth]{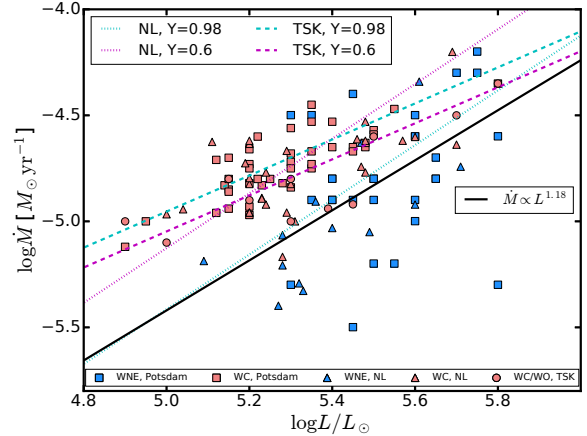}
\caption{Empirical mass-loss rates of hydrogen-free WNE, WC, and WO stars in our galaxy, compared 
with the NL and TSK prescriptions (dotted and dashed lines). The Potsdam, NL and TSK
samples are denoted by squares, triangles, and circles, respectively. WNE and WC/WO stars
are marked by blue and coral colors, respectively. 
Here, a correction for a clumping factor of $D = 10$ was applied to the mass-loss rates of the Potsdam WNE stars, 
to be consistent with the other empirical WR mass-loss rates (see the text). 
The thick black solid line gives
the result of our new prescription for WNE stars, based on the Potsdam WNE sample 
(Eq.~(3) with $f_\mathrm{WR} = 1.0$). \label{fig:rate}}
\end{center}
\end{figure}

The stellar wind theory suggests that WR winds are driven by radiation pressure
caused by metal lines~\citep[e.g.,][]{Graefener05, Vink05, Graefener08,
Puls08}.  This implies that WR loss rates should depend on the luminosity and
the chemical composition at the stellar surface.  The standard NL mass-loss
rate prescription is given by 
\begin{equation} \log \dot{M}_\mathrm{NL} = -11.0
+ 1.29\log\left(\frac{L}{{L_\odot}}\right) + 1.7\log Y + 0.5 \log Z~,
\end{equation} 
where $Y$ and $Z$ denote the surface mass fractions of helium
and metals.  The mass-loss rate $\dot{M}$ is given in units of
$M_\odot~\mathrm{yr}^{-1}$.  This is a very general prescription  that can be
applied for all types of WR stars including WNL, WNE, WC, and WO.  Note,
however, that this prescription is based on their selected sample of Galactic
WR stars. The metallicity dependence here is not related to the initial
metallicity but to the enrichment of carbon and oxygen at the surface due to
mass loss (i.e, $Z = 1-Y$ for hydrogen-free WR stars).  However, in many
stellar evolution models including those with the MESA code~\citep{Paxton11},
this $Z$ dependence is also used for considering the effect of the initial
metallicity.  

On the other hand, TSK suggests the following prescription based on their
selected sample of WC and WO stars in our galaxy, LMC, and IC1613:
\begin{equation}
\log \dot{M}_\mathrm{TSK} = -9.20 + 0.85\log\left(\frac{L}{{L_\odot}}\right) 
+ 0.44\log Y + 0.25 \log\left(\frac{Z_\mathrm{init}}{Z_\mathrm{\odot}}\right)~. 
\end{equation}
One of the advantages of the TSK prescription over the NL prescription is that
the dependence of the initial metallicity (or the iron metallicity $Z_\mathrm{Fe}$
as presented by TSK) is considered separately from the effect of
self-enrichment of CO elements.  This approach is consistent with the
theoretical studies that find different impacts of iron and CNO elements on
WR winds~\citep{Vink05, Graefener08}.

Fig.~\ref{fig:rate} presents the WR mass-loss rates given by NL and TSK at
solar metallicity, compared to the empirical values of Galactic hydrogen-free
WR stars\footnote{Some WNE stars are found to have small amounts of hydrogen
at their surfaces. However, only hydrogen-free WNE stars are considered in this
study}.  Note that the Potsdam group inferred mass-loss rates with clumping
factors of $D = 4$ for  Galactic WNE stars and with $D = 10$ for WC stars,
respectively~\citep{Hamann06, Sander12}.  Given that both NL and TSK
prescriptions are based on the data compatible with $D = 10$ rather than with
$D = 4$,  here we corrected the WNE mass-loss rates of the Potsdam group by a
factor of $(4/10)^{0.5}$ for this comparison.  This is because the empirical
mass-loss rates scale with $D^{-1/2}$.

\subsection{Dependence on the surface helium abundance}

\begin{figure}
\begin{center}
\includegraphics[width = \columnwidth]{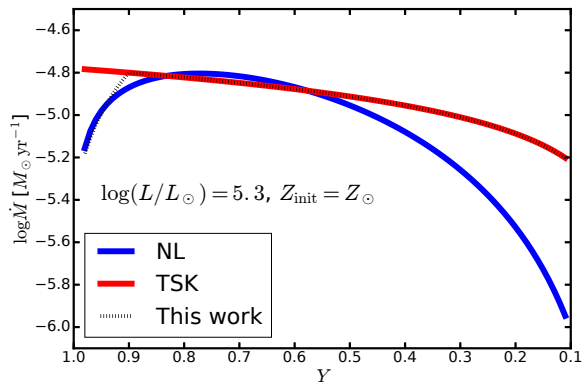}
\caption{
Mass-loss rates of hydrogen-free WR stars as a function 
of the surface helium mass fraction ($Y$) for $\log L/L_\odot = 5.3$ and $Z_\mathrm{init} = Z_\odot$
according to the prescriptions of NL (blue), TSK (red), 
and the present work with $f_\mathrm{WR} = 1.0$ (dotted line; Eqs.~(3), (4), \& (5)).
\label{fig:Ydep}}
\end{center}
\end{figure}

In Fig.~\ref{fig:rate},  it is clearly observed that the mass loss rates of  WC/WO
stars are systematically higher than those of WNE stars, for a given
luminosity.  The NL prescription gives a result consistent with this important
fact: in the figure, the mass-loss rate with $Y=0.6$ is higher by a factor of
2.52 than that with $Y = 0.98$, which results from the dependence of $\dot{M}
\propto Y^{1.7}(1-Y)^{0.5}$.  Fig.~\ref{fig:Ydep} illustrates this $Y$
dependence more clearly.  The NL rate increases rapidly during the transition
from WNE ($Y = 0.98$ for $Z_\mathrm{init} = Z_\odot$) to WC ($Y<0.98$) until
$Y$ reaches 0.77, after which the WC mass-loss rate gradually decreases. 

On the other hand, the TSK prescription does not properly reflect this systematic
difference between WNE and WC mass-loss rates.  Having the relation of $\dot{M}
\propto Y^{0.44}$, the TSK values with $Y=0.98$ is about 24\% higher than those with
$Y=0.6$, in contrast to the observation. This means that the TSK prescription,
which  is based on a WC/WO sample with $Y \le 0.975$, would result in a
significant overestimate when extrapolated to WNE stars.  TSK argues that their
prescription is compatible with the WNE mass-loss rates of the Potsdam group,
but they did not make a correction for $D=10$ to the Potsdam data in their
comparison (F. Tramper, private communication). 

On the other hand, both NL and TSK found that WC/WO stars with very low $Y$ have
systematically lower mass-loss rates than those with higher $Y$.  This fact is
reflected in both prescriptions as shown in Fig.~\ref{fig:Ydep}.  Note that the
$Y$ dependence of the TSK prescription is weaker than that of  NL.  TSK argues
that their prescription can better explain the mass-loss rates of WO stars with
$Y \lesssim 0.4$, for which the NL prescription gives too small values compared
to observations~\citep{Tramper15}. 

\subsection{Dependence on the luminosity}

The relation of $\dot{M} \propto L^{0.85}$ in the TSK prescription for WC/WO
stars is in good agreement with that inferred from the WC data of the Potsdam group,
which is $\dot{M} \propto L^{0.8}$~\citep{Sander12}. NL also gives a very
similar result of $\dot{M} \propto L^{0.84}$ with their WC
sample~\citep{Nugis00}.  As pointed out by TSK, therefore, the standard NL
prescription that comprises all WR types (WNL, WNE and WC/WO) has a too steep
dependence on the luminosity ($\dot{M} \propto L^{1.29}$) compared to the empirical
result with WC stars. 

For WNE stars, the NL sample gives $\dot{M} \propto L^{1.27}$, which is much
steeper than that of WC/WO stars.  The luminosity dependence is less clearly found
if we include the Potsdam WNE sample. The large scatter of the WNE mass-loss rates
in Fig.~\ref{fig:rate} is likely due to the uncertainty of the distance
measurement. \citet{Hainich14} found $\dot{M} \propto  L^{1.18}$ for single WNE
stars in the LMC, for which the distance uncertainty is much smaller than in the
case of Galactic WR stars. Therefore, it seems to be real that WNE stars have a
steeper luminosity dependence of mass-loss rates compared to the case of WC/WO
stars.

\subsection{Dependence on the initial metallicity}\label{sect:initialZ} 

As mentioned above, TSK found an initial metallicity dependence of $\dot{M}
\propto Z_\mathrm{init}^{0.25}$ for WC/WO mass-loss rates.  \citet{Hainich14}
found a much steeper relation of $\dot{M} \propto Z_\mathrm{init}^{0.9}$ for WN stars
including WNL. This empirical results are in qualitative agreement with the
theoretical study by \citet{Vink05}, who found a steeper initial metallicity
dependence for WNL stars than for WC stars.

Using the Potsdam data of \citet{Hamann06} and
\citet{Hainich14}, we may also derive an initial metallicity dependence only
for WNE stars.  For this purpose, we may use the relation of $\dot{M} \propto
L^{1.18}$ that is found with single WNE stars in the LMC~\citep{Hainich14}.  Given
that the scatter in the distance estimate  is minimized with the LMC sample, we
may consider this relation more reliable than that obtained with the Galactic
sample. Using the fixed exponent of 1.18, the mass loss rates of Galactic WNE
stars of the Potsdam sample can be fitted to $\log \dot{M} = -11.32 + 1.18\log
(L/{L_\odot})$ with a mean squared error of 0.15.

This relation gives 51 per cent higher mass-loss rates than those of LMC
WNE stars (i.e., $\log \dot{M} = -11.5 + 1.18\log(L/{L_\odot})$;
\citealt{Hainich14}). 
Assuming that LMC metallicity is half the solar value as in TSK,
we get the metallicity dependence of  $\dot{M} \propto Z_\mathrm{init}^{0.6}$.
\footnote{TSK suggested $\dot{M} \propto
(Z_\mathrm{init}/Z_\mathrm{\odot})^{1.3}$ for WNE stars using the same data we
used.  This discrepancy is because they did not correct the Galactic WNE
mass-loss rates for $D=10$, as mentioned above.  \citet{Hainich14} used $D=10$
in their analysis of LMC WNE stars and a correction of $(4/10)^{0.5}$ to
Galactic WNE mass loss rates by \citet{Hamann06} is needed to derive the
metallicity dependence in a consistent way.}
This is less steep than in the case where  WNL stars are included (i.e., $\dot{M} \propto
Z^{0.9}_\mathrm{init}$) but still
much steeper than in the case of WC/WO stars (i.e., $\dot{M} \propto
Z^{0.25}_\mathrm{init}$). This fact should be taken into
account when addressing the metallicity effect of WR star evolution.

\subsection{Towards a better prescription for mass-loss rates of hydrogen-free WR stars}

The above discussion on previous studies on hydrogen-free WR mass-loss rates, which are based on the best data
currently available,  leads to the following conclusions. 
\begin{enumerate}
\item WC/WO stars have systematically higher mass-loss rates than WNE stars. 
\item Among WC/WO stars, mass-loss rates tend to decrease as the surface abundance of helium decreases.
\item  The mass-loss rates of WNE and WC/WO stars have different luminosity dependencies, with those of WC/WO stars having a less steep
dependence than those of WNE stars. 
\item The dependence of WNE mass-loss rates on the initial metallicity is steeper than that of WC/WO stars. For WC/WO stars, the effects of the initial metallicity and the self-enrichment of carbon and oxygen on mass loss rates should be considered separately.   
\end{enumerate}

A good prescription for WR mass-loss rates should take into account all the
above facts.  The mass-loss rate prescriptions for hydrogen-free WR stars
suggested in the 80s and 90s~\citep[e.g.,][]{Langer89, Vanbeveren98,
Wellstein99} only consider a mass or luminosity dependence, and none of the
above conditions are satisfied. In many recent stellar evolution models including
those with the BEC and MESA codes~\citep{Brott11, Paxton11}, the standard NL
prescription given by Eq.~(1) that combines WN and WC data is used.  This
standard NL prescription meets the first and second requirements, but does not
properly consider the third one, having a single luminosity dependence for all
WR types (i.e, $\dot{M} \propto L^{1.29}$). 
Some authors use  two different prescriptions for WN and WC stars presented by NL
instead of the standard NL prescription that combines the two \citep[e.g.,][]{Eldridge06}.  
In this case, the first three conditions can be satisfied. However, 
NL only analysed  Galactic WR stars and did not address the dependence on the initial metallicty.  
The TSK prescription given by Eq.~(2) is only suitable for WC/WO stars.

We therefore suggest the following approach. For WNE stars, we may 
use the empirical relation obtained with the Potsdam data (see Sect.~\ref{sect:initialZ}): 
\begin{equation}
\dot{M}_\mathrm{WNE} = f_\mathrm{WR} \left(\frac{L}{{L_\odot}}\right)^{1.18}\left(\frac{Z_\mathrm{init}}{Z_\mathrm{\odot}}\right)^{0.60} 10^{-11.32} ~~\text{for}~~Y = 1 - Z_\mathrm{init}~~. 
\end{equation}
Note that the
3rd term $(Z_\mathrm{init}/Z_\mathrm{\odot})^{0.60}$ may be replaced
by the iron metallicity (i.e., $(Z_\mathrm{Fe}/Z_\mathrm{Fe,\odot})^{0.60}$). 
Here, we also introduce a scaling factor $f_\mathrm{WR}$, to consider the
uncertainty in the empirical estimates of WR mass-loss rates. For example, if
$D=4$ was adopted instead of $D=10$, the mass-loss rates presented in
Fig.~\ref{fig:rate} would increase by 58\%.  We propose below that this scaling
factor should be calibrated with Galactic WR stars. 

For WC/WO stars, we suggest using the TSK prescription.  This is because it
provides an improved description of the mass-loss rate dependencies on the surface helium 
abundance and the initial metallicity as discussed in their paper.  Its luminosity dependence is also
consistent with the results of the other groups (NL and Potsdam) as mentioned above:
\begin{equation} 
\dot{M}_\mathrm{WC}
= f_\mathrm{WR} \dot{M}_\mathrm{TSK}~~ \text{for}~~ $Y < 0.90$~. 
\end{equation}
Here we assume that $Y < 0.90$ for WC stars, for which nitrogen completely disappears from the surface. 
In principle, WNE and WC/WO stars might have different values of $f_\mathrm{WR}$ if clumping properties depended
on the spectral types of WR stars. In this study, however, we just take the simplest assumption
that $f_\mathrm{WR}$ is  the same for all WR types given that the clumping physics is still not well understood.

For $0.9 \le Y < 1 - Z_\mathrm{init}$, we suggest using an
interpolated value between $\dot{M}_\mathrm{WNE}$ and $\dot{M}_\mathrm{WC}$, 
to consider the enhancement of the mass-loss rate during the transition phase from WNE to WC, as the following:
\begin{equation}
\dot{M}_\mathrm{TR} = (1-x)\dot{M}_\mathrm{WNE} + x\dot{M}_\mathrm{WC}~~\text{for}~~ 0.90 \le Y < 1 - Z_\mathrm{init}~~, 
\end{equation}
where $x$ is given by 
\begin{equation}
x = (1 - Z_\mathrm{init} - Y)/(1 - Z_\mathrm{init} - 0.9)~~. 
\end{equation}

In this way, all the four requirements for the mass-loss rate prescription for
hydrogen-free WR stars can be fulfilled.  In Fig.~\ref{fig:Ydep}  the mass-loss
rate according to our prescription is compared to those given by the NL and TSK
prescriptions.  As expected, our prescription gives a value comparable to the
NL rate when $Y$ is close to 0.98 and follows the TSK rate for $Y < 0.9$. 

\section{Evolutionary models v.s. Observed WR stars}\label{sect:evol}

\begin{figure*}
\begin{center}
\includegraphics[width = 0.45\textwidth]{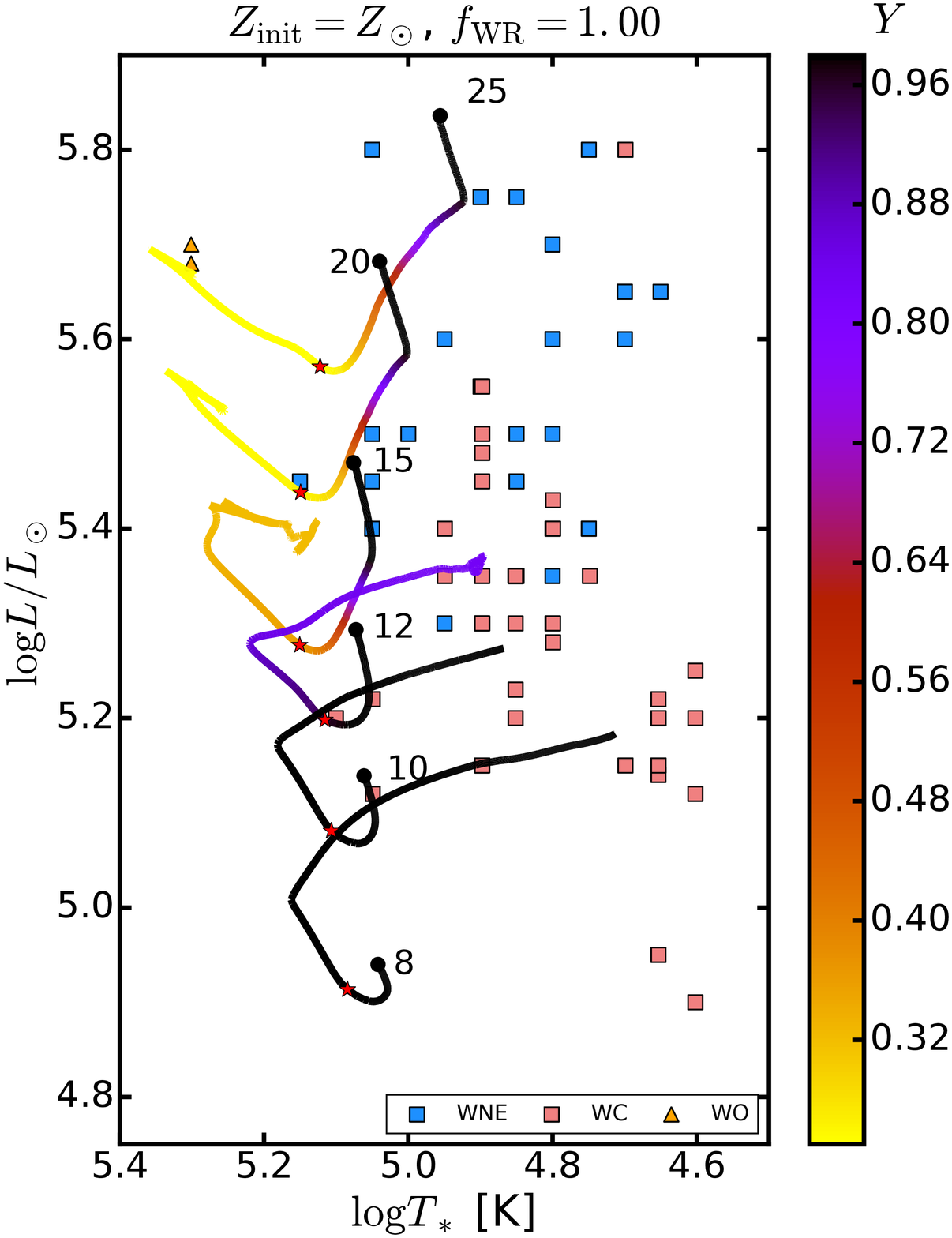}
\includegraphics[width = 0.45\textwidth]{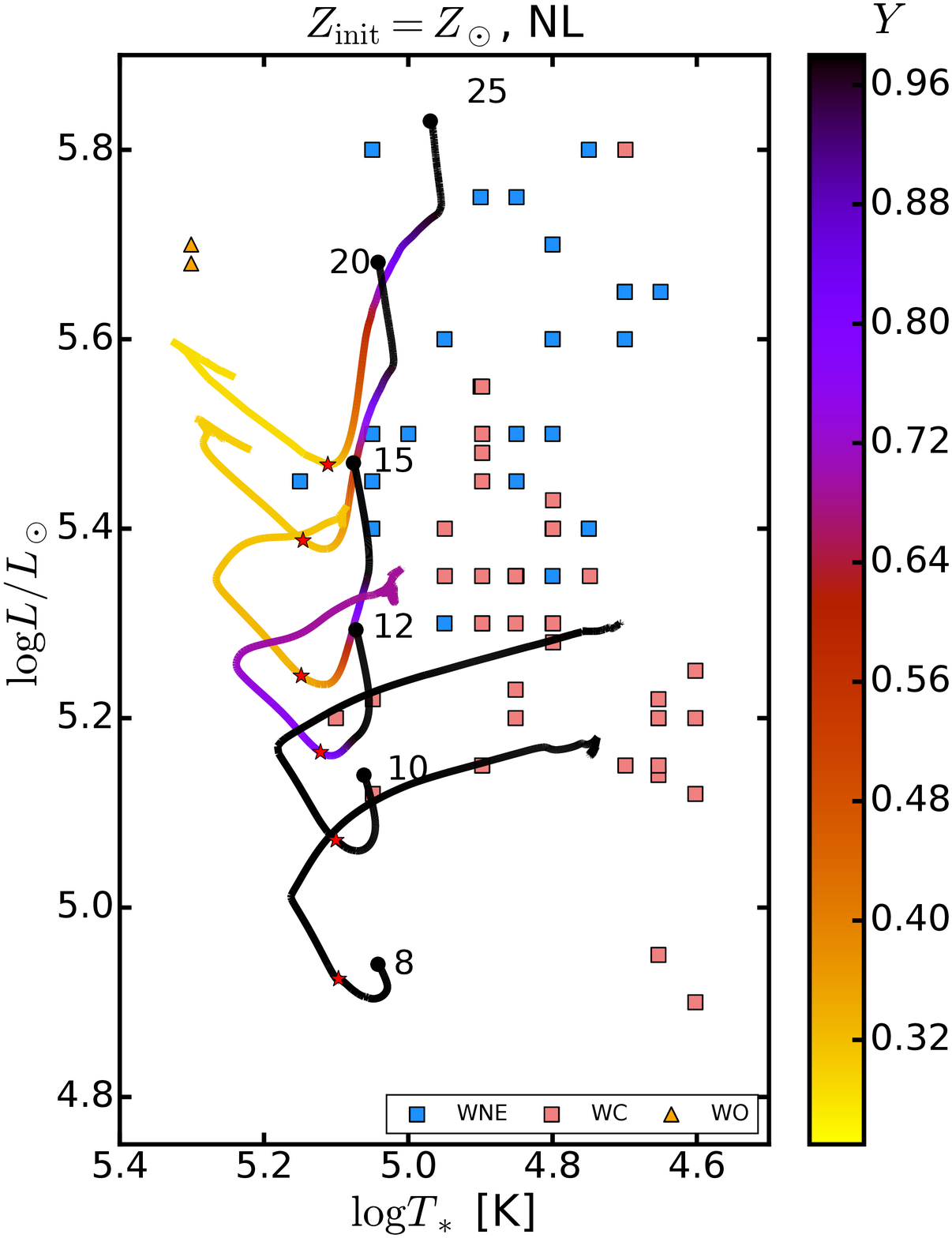}
\includegraphics[width = 0.45\textwidth]{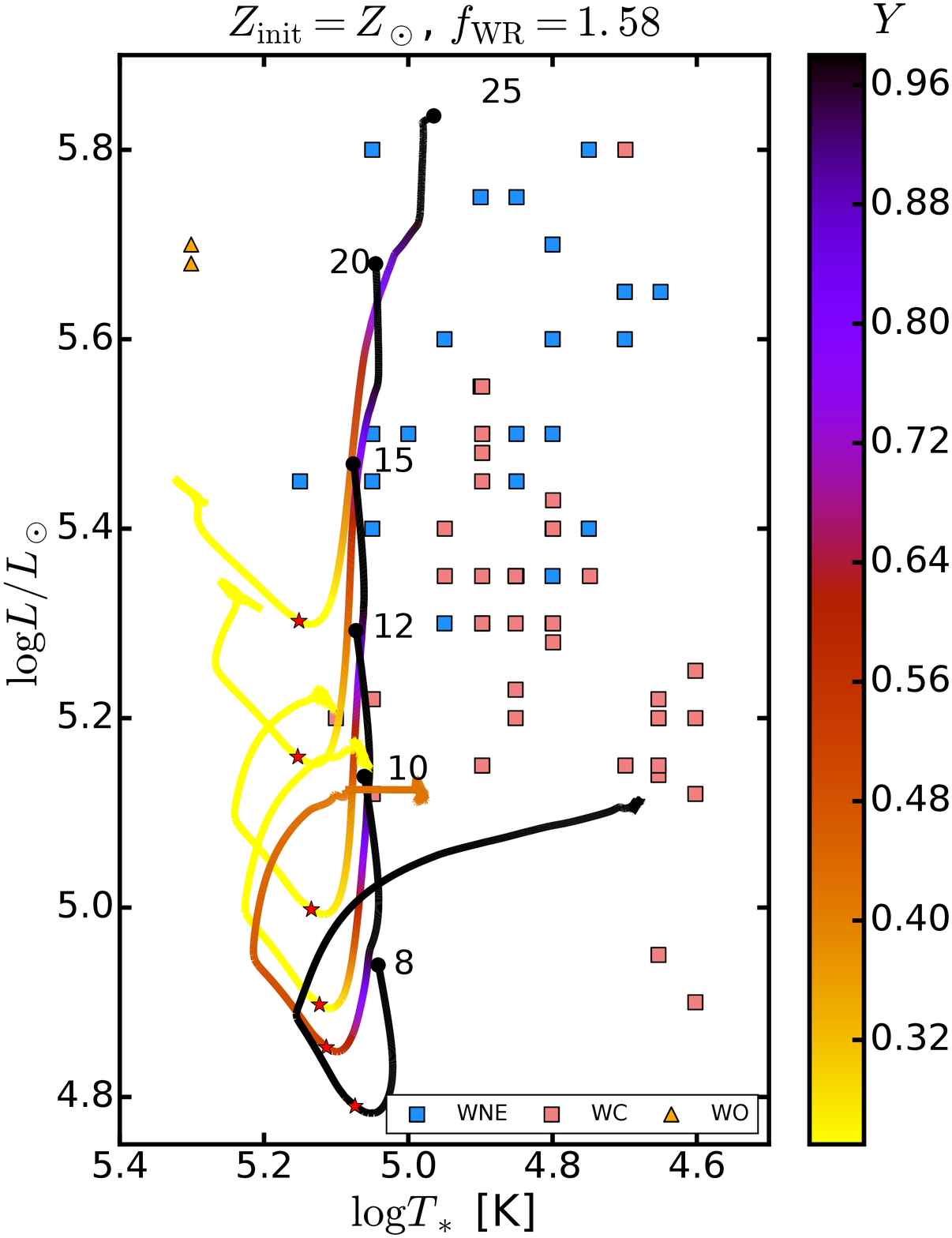}
\includegraphics[width = 0.45\textwidth]{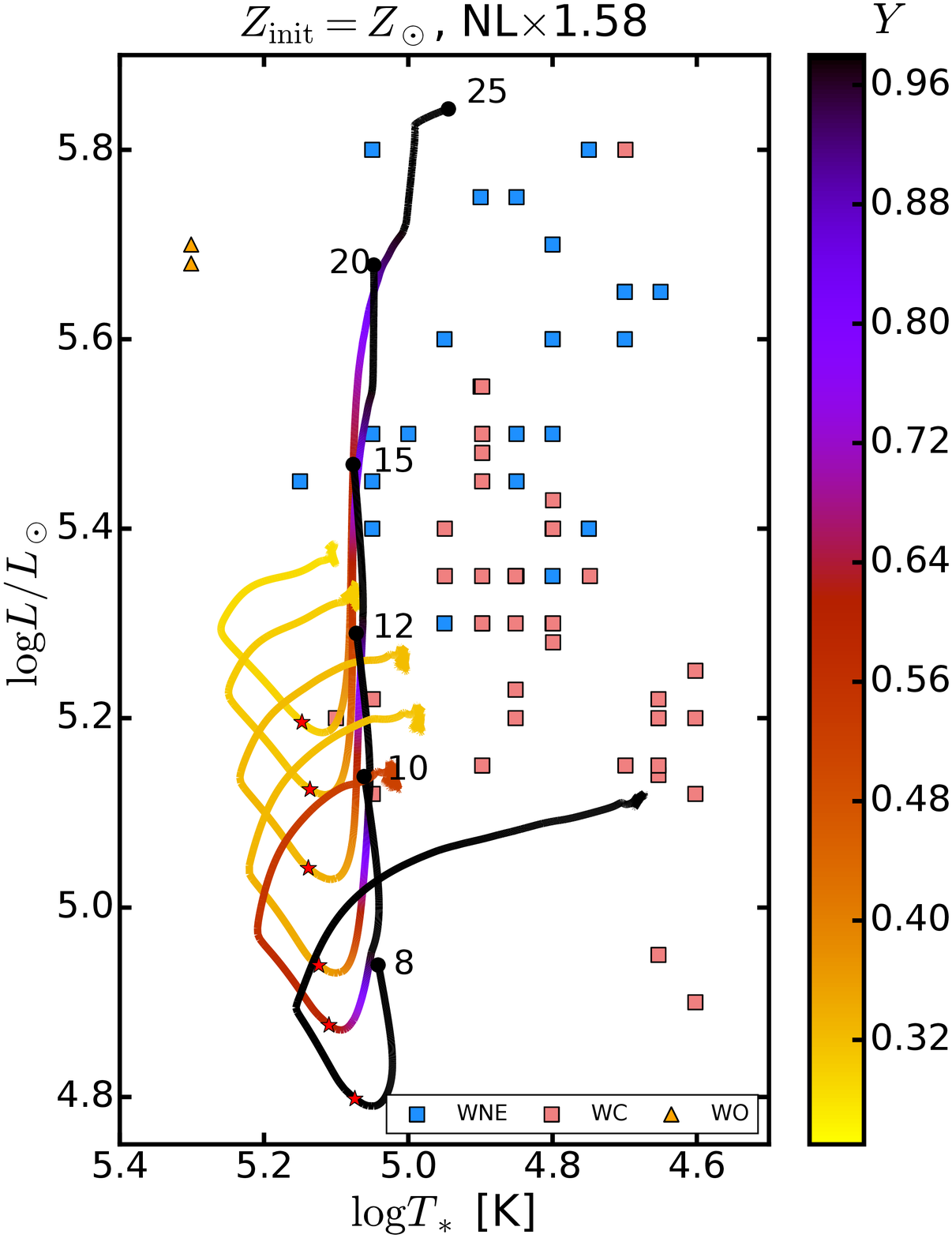}
\caption{Evolutionary tracks of He stars at solar metallicity on the HR diagram with
our mass-loss rate prescription using $f_\mathrm{WR} = 1.0$ (top left) and $f_\mathrm{WR} = 1.58$ (bottom left), 
and with the NL prescription (top right) and the NL prescription multiplied by 1.58 (bottom right). The color shading 
on each track and the color bar on the right-hand side of each panel indicate
the surface mass fraction of helium~($Y$). 
The black filled circle and red star symbol on each track mark 
the beginning of core He burning and core helium exhaustion, respectively. 
The blue and  coral squares
respectively denote  the Galactic WNE and WC stars in the Potsdam sample that are not in binary systems~\citep{Hamann06, Sander12}. 
The orange triangles denote the Galactic WO stars in the Potsdam sample~\citep{Sander12}.  
\label{fig:hr}}
\end{center}
\end{figure*}

\begin{figure} 
\begin{center} 
\includegraphics[width = 0.98\columnwidth]{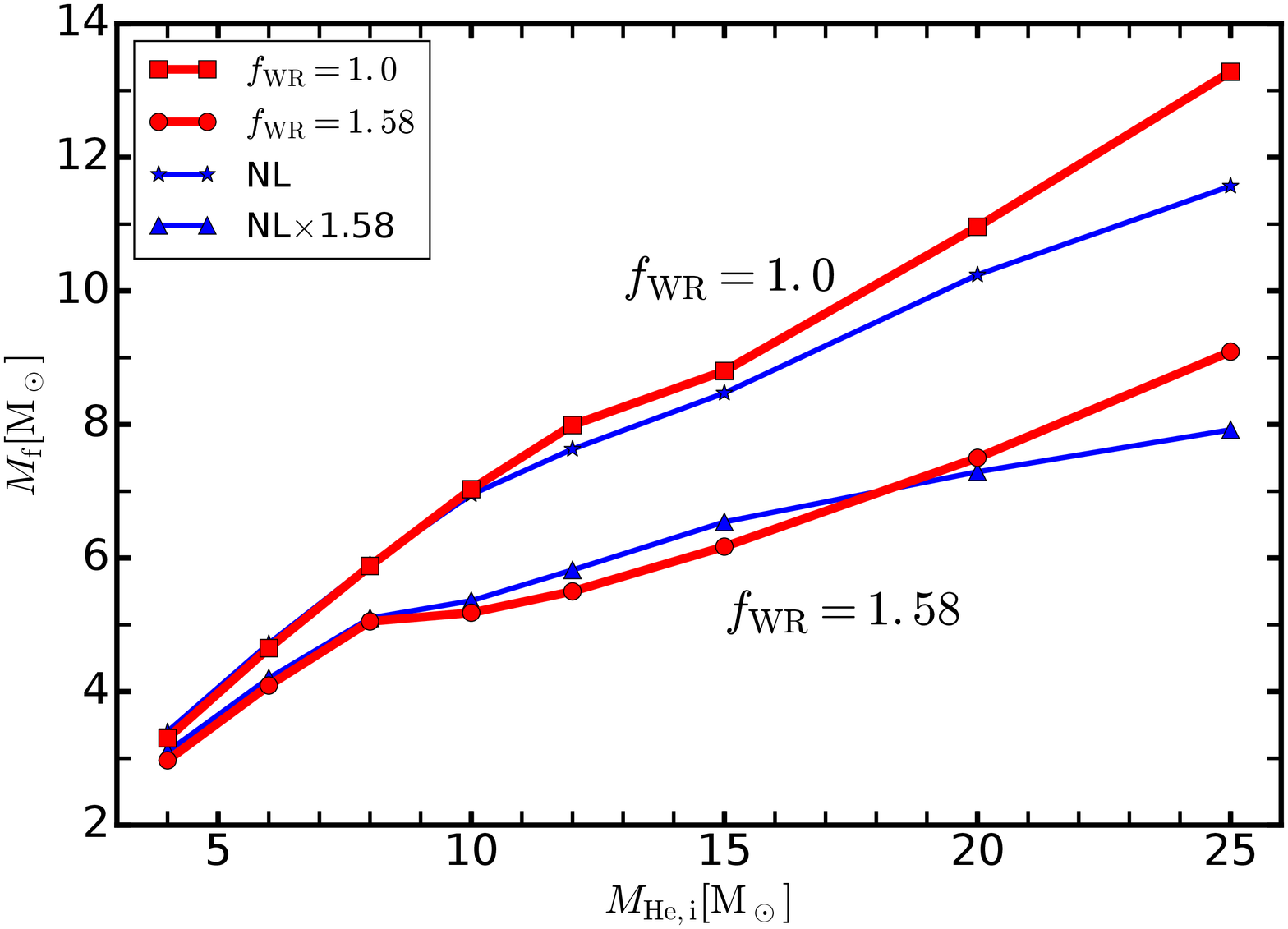} 
\caption{Final masses of He star models at solar metallicity as a function of the intial mass. 
The red connecting lines give the results with our mass-loss rate prescription
using $f_\mathrm{WR} = 1.0$ (square) and $f_\mathrm{WR}=1.58$ (circle).   
The results with the NL prescription (star) and 1.58$\times$ NL prescription (triangle)
are given by the blue connecting lines. 
\label{fig:finalmass}} 
\end{center}
\end{figure}

\begin{figure}
\begin{center}
\includegraphics[width = 0.98\columnwidth]{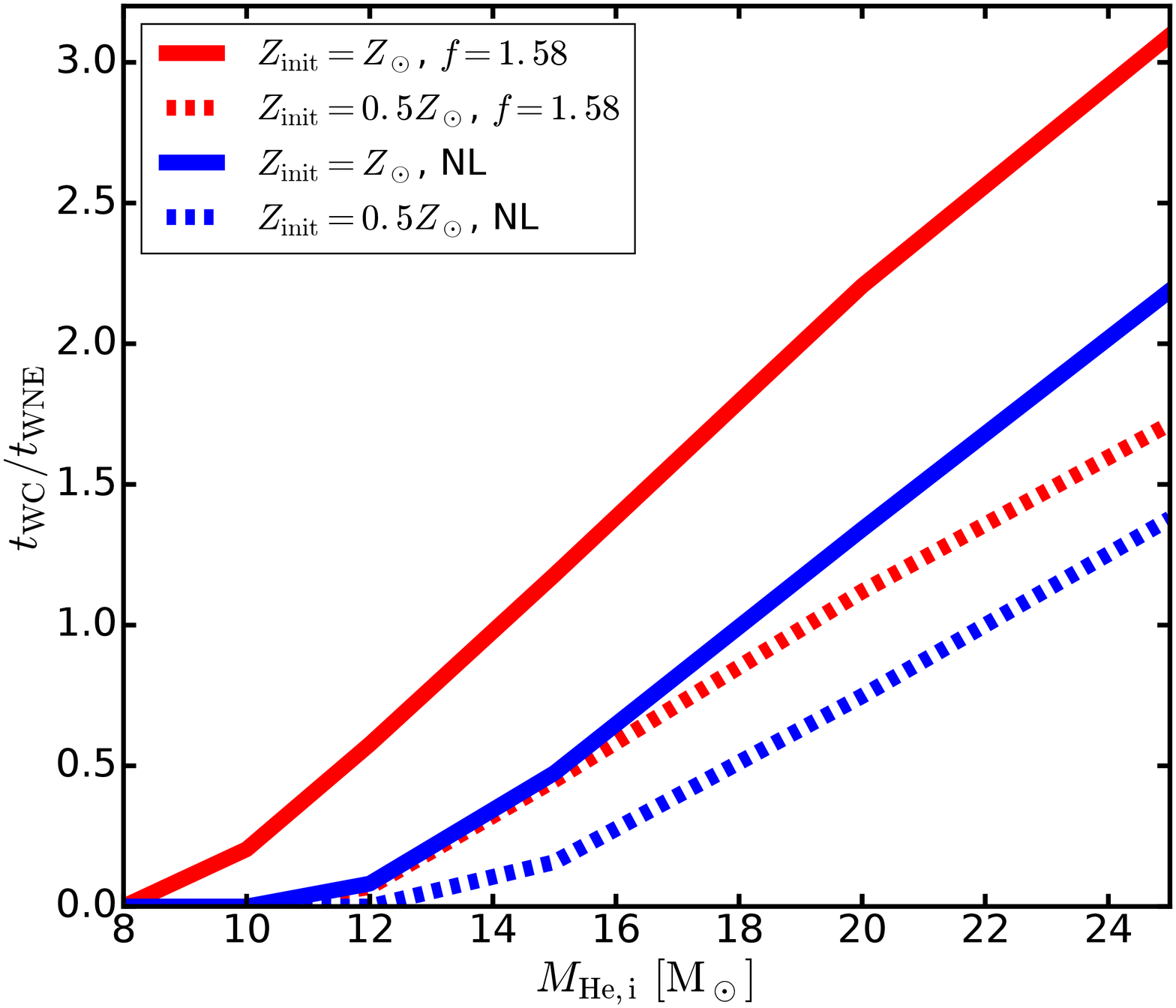}
\caption{The ratio of  WC to WNE lifetimes as a function 
of the initial mass of the He star. The red lines
denote the results with our prescription using $f_\mathrm{WR} = 1.58$
at solar (solid line) and LMC (dashed line) metallicities. 
The blue lines are the corresponding results with the standard NL prescription~\label{fig:tratio}}
\end{center}
\end{figure}

Here we present stellar evolutionary models of pure He stars using our new
prescription for WR mass-loss rates. For comparison, models with the NL
prescription are also presented.   We used the BEC code for the calculations,
which has been widely used for massive stars~\citep[see][and references
therein]{Yoon10, Brott11}. The Schwarzschild criterion for convection was adopted, and
$Z_\mathrm{init} = 0.02$ and $Z_\mathrm{init}=0.01$ were chosen for
representing solar and LMC metallicities.  All the model sequences were
calculated up to core neon burning or beyond.  The considered initial masses of
He stars are 4, 6, 8, 10, 12, 15, 20 and 25~${M_\odot}$.  The results of the
calculations are summarized in Tables~\ref{tab1} and~\ref{tab2}.  

\subsection{Solar metallicity models with $f_\mathrm{WR} = 1.0$} 

The evolutionary tracks of solar metallicity models on the Hertzsprung-Russel
(HR) diagram are shown in Fig.~\ref{fig:hr}.  Let us first discuss the models
with $f_\mathrm{WR} = 1.0$. 

We notice some minor differences between the results with our new prescription
and those with the NL prescription. The NL prescription gives somewhat higher
mass loss rates for  WNE stars of $\log L/{L_\odot} \gtrsim 5.0$ and for WC
stars of $\log L/{L_\odot} \gtrsim 5.2$ than our
prescription~(Fig.~\ref{fig:rate}).  The relatively massive He stars
($M_\mathrm{He,i} \ge 10~{M_\odot}$) lose more mass with the NL prescription,
as a result (Fig.~\ref{fig:finalmass}). For example, the final mass of the
25~${M_\odot}$ He star is 11.57~${M_\odot}$ with the NL prescription and
13.28~${M_\odot}$ with  our prescription, respectively.  For lower initial
masses ($M_\mathrm{He,i} \le 8.0~{M_\odot}$), our prescription gives slightly
higher mass-loss rates and the resultant final masses are lower, although the
difference is very small. 

Comparison with the Galactic WR stars reveal that none of the NL and our
prescription can properly explain the relatively faint WC stars ($4.9 \lesssim
\log L/{L_\odot} \lesssim 5.3$), of which the luminosities are lower than the
faintest WNE star ($\log L/{L_\odot} \approx 5.3$).  The 12~${M_\odot}$ He star
can make a WC star of which the luminosity can become as low as $\log
L/{L_\odot} = 5.16$ with the NL prescription,  but the WC lifetime at this low
luminosity is very short ($\sim 10^{4}$~yr; Table~\ref{tab1};
Fig.~\ref{fig:tratio}). None of the two prescriptions can produce a WC star
with $\log L/L_\odot < 5.16$. Therefore, the large number fraction of WC stars
of this low luminosity range cannot be explained with these models. 

This discrepancy between the theoretical prediction and observation has already
been noted by \citet{Sander12}. The Geneva group models~\citep{Meynet05,
Georgy12} where the standard mass-loss rate prescriptions for red supergiant
stars and WR stars (i.e., de Jager and NL rates, respectively) were used fail to
reproduce the observed luminosity range of WN and WC stars.  In particular, WC
stars are predicted to be systematically much more luminous ($\log
L/{L_\odot} \gtrsim 5.4$)  than Galactic WC stars in these studies. 

Note that  this problem cannot be solved by binarity. He stars of 8 --
12~${M_\odot}$ having $\log L/{L_\odot}\simeq 4.9 - 5.3$ may be produced to
become  WN stars by binary interactions. But the subsequent evolution of these
WN stars is largely determined by WR wind mass loss like single
stars~\citep[e.g.,][]{Yoon10, Eldridge13, Yoon17}, and  whether He stars are in
binary systems cannot significantly affect our discussion.  Furthermore, there
is no apparent evidence for binarity with the WR sample in the figure, for which
we excluded WR stars identified in binary systems. 

In conclusion, the standard NL prescription gives too low mass-loss rates to
explain relatively faint WC stars ($\log L/{L_\odot} \lesssim 5.3$).  Revising the
luminosity and chemical composition dependencies of the WR mass-loss rate does
not improve the situation.  Our new mass-loss rate prescription has only a
minor effect on the evolution on the HR diagram, compared to the result with
the NL prescription, although it has significant impact on SN Ib/Ic progenitor
models (see Sect.~\ref{sect:snIbc} below).  


\subsection{Solar metallicity models with $f_\mathrm{WR} = 1.58$}

Although great progress has been made in radiative transfer modelling of WR
star atmospheres during the last two decades, the inferred mass-loss rates of
WR stars still depend on some free parameters, including the wind velocity law
and clumping factor~\citep{Puls08}.  In particular, the mass-loss rate scales
with the clumping factor $D$ as $\dot{M} \propto D^{-1/2}$.  Several studies
prefer $D = 10$~\citep[e.g.,][]{Sander12, Hainich14}, but it is not clear yet
if the complex features resulting from hydrodynamic instabilities in the wind
material can be well described by this single free parameter. This allows us to
have some freedom in our choice of $f_\mathrm{WR}$.  

We get $f_\mathrm{WR} = 1.58$ if we adopt $D=4$ following \citet{Hamann06}.
In the lower panel of Fig.~\ref{fig:hr}, evolutionary tracks of solar
metallicity He star models with $f_\mathrm{WR} = 1.58$ are presented.  

One of the most striking features here is the downward evolution of He stars almost
vertically on the HR diagram  during the core helium burning phase.  The dynamic
range of the luminosity of a He star for a given initial mass is significantly
enlarged (by more than a factor of 2) compared
to the case of $f_\mathrm{WR} = 1.0$.  
For example, the 15~${M_\odot}$
He star  covers relatively wide luminosity ranges of $5.32 \lesssim \log
L/{L_\odot} \lesssim 5.47$  and  $5.00 \lesssim \log L/{L_\odot}
\lesssim 5.32$ for WNE and WC, respectively. These
ranges are reduced to $5.37 \lesssim \log L/{L_\odot} \lesssim 5.47$ for
WNE and $5.28 \lesssim \log L/{L_\odot} \lesssim 5.37$ for WC, with
$f_\mathrm{WR} = 1.0$. 

Interestingly, with $f_\mathrm{WR} = 1.58$, the 15~$M_\odot$ He star model can
roughly explain the lower limits of both WNE and WC star luminosities in the
Potsdam sample ($\log L/{L_\odot} \approx$ 5.3 and 4.9, respectively).  The
zero-age main sequence (ZAMS) mass that corresponds to a 15 He star is about 30
-- 40~${M_\odot}$ depending on the degree of core overshooting on the main
sequence. This ZAMS mass is high enough for a single star to lose its entire
hydrogen envelope during the RSG phase~\citep{Meynet05, Georgy12}.  Note also
that, with $f_\mathrm{WR} = 1.58$,  the lifetimes of WC stars can be longer
than those of WNE for $M\gtrsim 14~{M_\odot}$ as shown in
Fig.~\ref{fig:tratio}. This means that the large number of relatively faint WC
stars ($\log L/L_\odot < 5.3$) may be explained by the evolution of He stars of
$\sim 15~M_\odot$. 

This points to the need of higher WR mass-loss rates than given by the standard
NL prescription to better explain observations.  More specifically, based on
our result with $f_\mathrm{WR} = 1.58$,  the observed WR population may be
explained by the following scenario that is fully consistent with the
Conti scenario.  WNE stars of $M \gtrsim 14 - 15~{M_\odot}$ ($\log L/L_\odot \gtrsim 5.3$) are
produced from single stars of $M > \sim 30~{M_\odot}$
via mass loss due to RSG winds, while lower ZAMS mass would not lead to
formation of WR stars.  WC stars are produced from these WNE stars by further
mass loss during the core He-burning phase.  This scenario can explain 1) the
reason why WC stars are systematically less luminous than WN stars, and 2) the
paucity of WNE stars and the large number of  WC stars in the luminosity range
of $4.9 \lesssim \log L/{L_\odot} \lesssim 5.3$.

\citet{Sander12}  noted that models of \citet{Vanbeveren98} can better explain
the luminosity distribution of WC stars than those of the Geneva group.  The
mass-loss rate prescriptions adopted by Vanbeveren et al. provide much higher values
for both RSG and WR stars than given by the commonly used prescriptions of
\citet{deJager88} and NL. This supports our suggestion of increasing the WR
mass-loss rate.  However, the WR mass-loss rate prescription of Vanbeveren et al.
(i.e., $\dot{M}_\mathrm{WR} = -10 + \log L/L_\odot$) does not consider the
difference between WN and WC stars, which is significant as discussed in
Sect.~\ref{sect:wind}.  

In contrast, \citet{Vanbeveren07} argued for the need of a decrease in  WR
mass-loss rates by a factor of two compared to those given by the NL
prescription.  This conclusion resulted from their comparison of the WC to WN
number ratio between the prediction of their binary star population models and
observations. But the WC sample by \citet{Sander12} was not available at that
time. A decrease in WR mass-loss rates would make the problem of the luminosity
discrepancy between models and observations even worse.  Note also that the
prediction of the WC/WN ratio sensitively depends on many uncertain physical
processes including overshooting, rotation, RSG mass loss, and binary
interactions~\citep{Eldridge08}.  The mass-loss rates of WNL stars would also
play an important role here, which is not addressed in the present study.
Therefore, we cannot make a detailed prediction on this with our new He star models
because it requires full stellar evolution models from the zero-age main
sequence (ZAMS). This should be a subject of future study.

In the initial and final mass diagram (Fig.~\ref{fig:finalmass}), there is a
notable change from $M_\mathrm{He,i} > 8.0~{M_\odot}$ in the slope of
the line with $f_\mathrm{WR} = 1.58$.   This results from the fact
that He stars  of $M_\mathrm{He,i} \gtrsim 10~M_\odot$ undergo the WC phase 
where WR mass-loss rates are significantly enhanced compared to those of WN
stars  as discussed in Sect.~\ref{sect:wind}.  Compared to the result with the
NL rate multiplied by 1.58~(Table~\ref{tab1}), our prescription with
$f_\mathrm{WR} = 1.58$ results in  lower final masses and surface He mass
fractions for $10 \le M_\mathrm{He,i}/{M_\odot} \le 15$, because our
prescription gives significantly higher mass-loss rates than the NL
prescription for relatively faint WC stars ($\log L/{L_\odot} < 5.2$;
Fig.~\ref{fig:rate}), for a given scaling factor $f_\mathrm{WR}$. 

For $M_\mathrm{He,i} \ge 20~{M_\odot}$,  in contrast, our prescription leads to
lower mass-loss rates in general and higher final masses than  in the case of
the 1.58$\times$NL prescription.  The surface helium mass fractions are lower in these
massive He star models because of the weaker dependence of the TSK prescription
on $Y$, which we use for WC/WO stars.  This leads to more stripping of the
helium envelope during the final evolutionary stages where $Y \lesssim 0.5$.
Note also that the final mass of the 25~$M_\odot$ He star is 9.1~$M_\odot$,
which would be massive enough to form a BH.   This implies that He stars of
$M_\mathrm{He,i} \gtrsim 25~M_\odot$ can produce a black hole (BH), and the
problem with BHs in the old days (i.e, the difficulty in explaining stellar mass BHs
with $M \gtrsim 7~M_\odot$ in BH binary systems at solar metallicity because of
too high WR mass-loss rates; e.g., \citealt{Nelemans01}) would not occur with
our moderate increase of WR mass-loss rates.  We discuss implications of this
result for SN Ib/Ic progenitors in Sect.~\ref{sect:snIbc} below. 

\begin{figure*}
\begin{center}
\includegraphics[width = 0.45\textwidth]{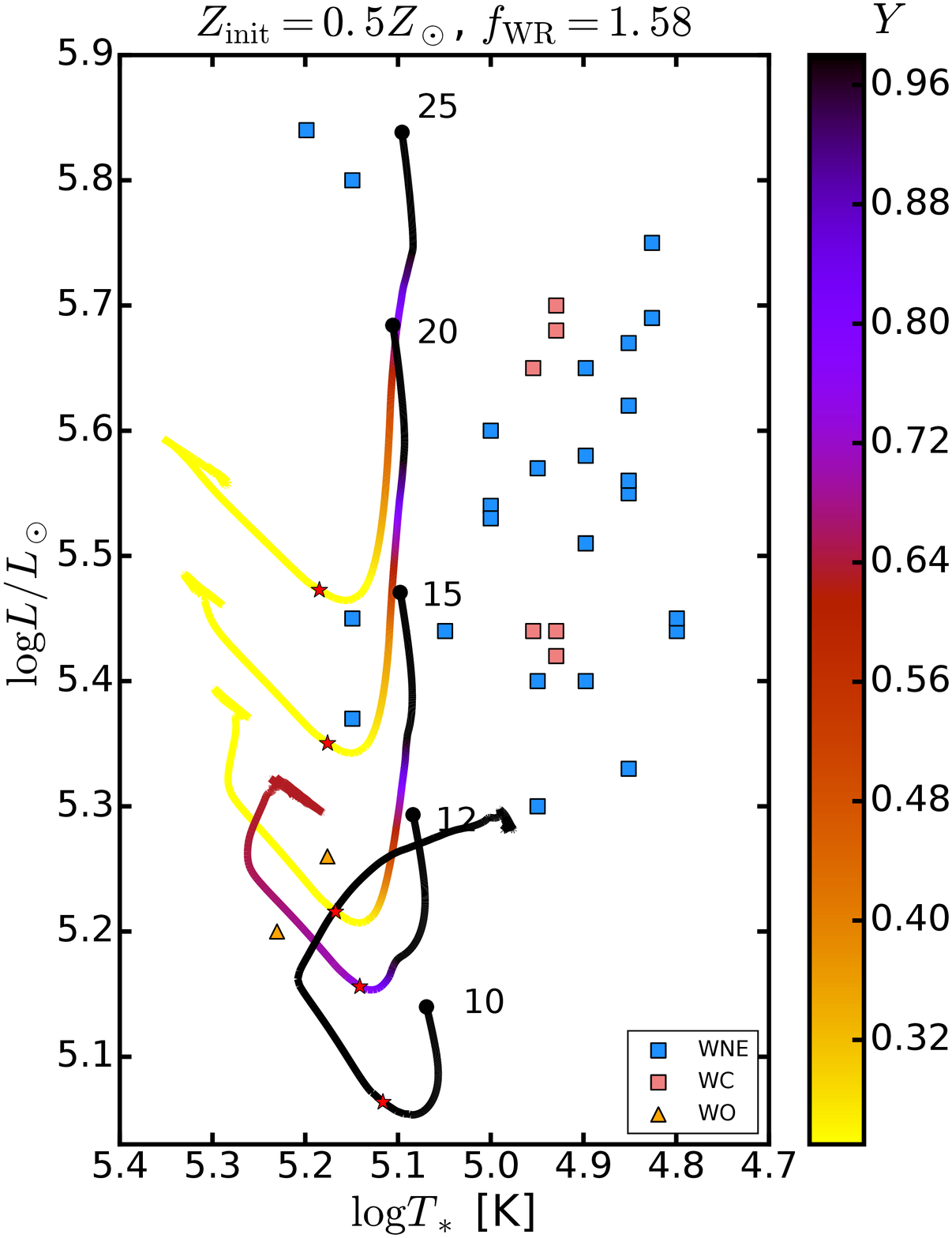}
\includegraphics[width = 0.45\textwidth]{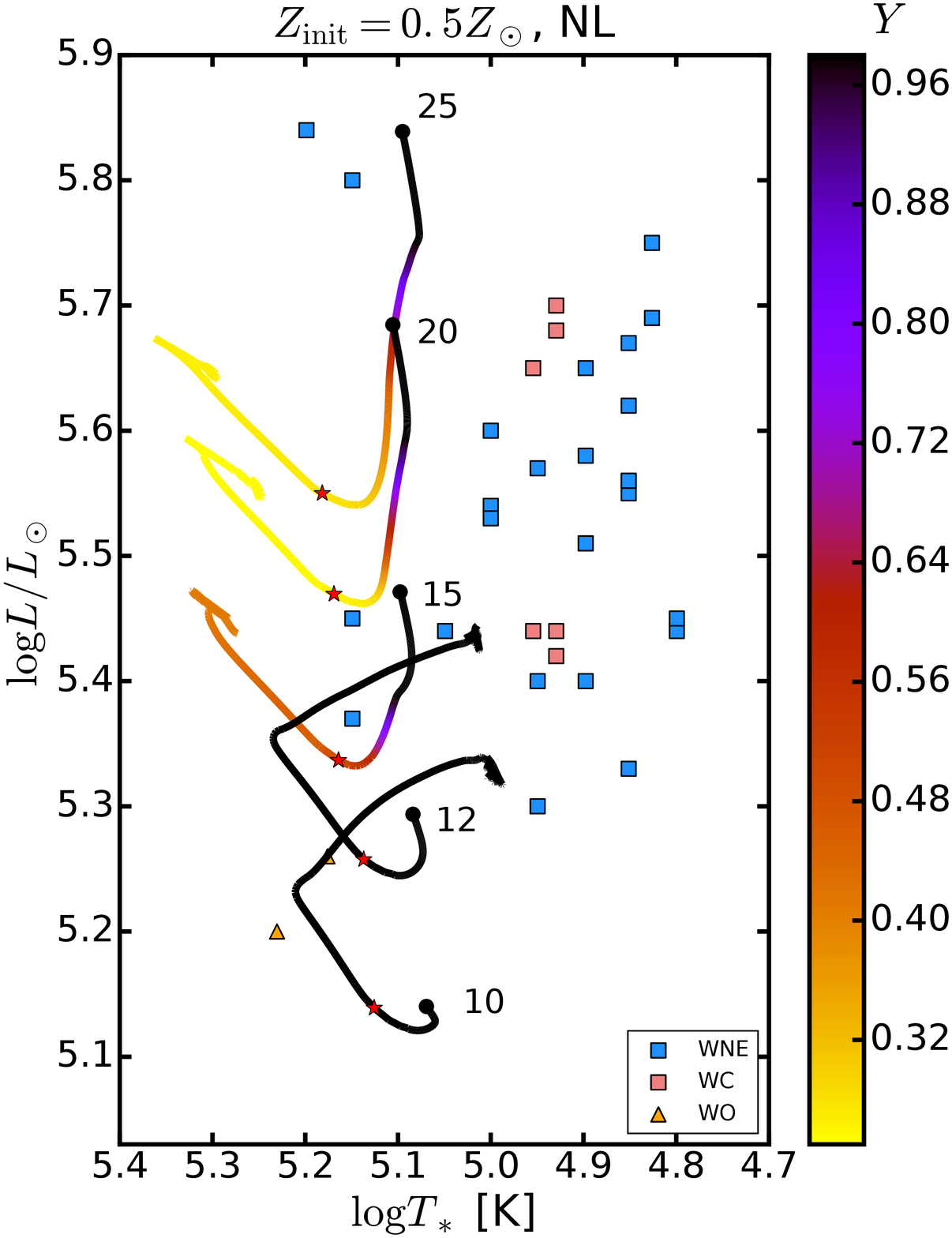}
\caption{Same with Fig.~\ref{fig:hr} but for LMC metallicity. The blue squares
denote the LMC WNE stars in the Potsdam sample~\citep{Hainich14}, 
and the  coral squares are the LMC WC stars of \citet{Crowther02}.
The two orange triangles are the LMC WO stars in the TSK sample. 
The left and right panels give the results with our mass-loss rate prescription 
using $f_\mathrm{WR} = 1.58$ and with the NL prescription, respectively.
\label{fig:hrlmc}}
\end{center}
\end{figure*}

\subsection{LMC metallicity models with $f_\mathrm{WR} = 1.58$}

LMC WR stars of WNE, WC and WO types on the HR diagram are shown in
Fig.~\ref{fig:hrlmc}.  Although WC stars of the Crowther
sample~\citep{Crowther02} have fairly high luminosities ($\log L/{L_\odot} >
5.4$), the two WO stars are  very faint ($\log L/{L_\odot} \simeq 5.20 - 5.25$)
compared to other WR stars in the LMC.  Like in the case of Galactic WR stars, the
standard NL prescription cannot explain these faint WO stars because He star
models of this luminosity does not become He-poor at LMC metallicity.  When
using our prescription with $f_\mathrm{WR} = 1.58$,  this luminosity limit is
predicted to be $\log L/{L_\odot} \approx 5.2$, which agrees well with the
observation.  

In conclusion, an overall increase in WR mass loss rates may well explain the
lower luminosity limits of WC/WO stars in both Galactic and LMC WR stars.  As
discussed below, this also leads to SN Ib/c progenitor models that are more
consistent with observations than those with the NL prescription.

\section{Implications for SN Ib/Ic progenitors}\label{sect:snIbc}

\begin{figure}
\begin{center}
\includegraphics[width = 0.48\textwidth]{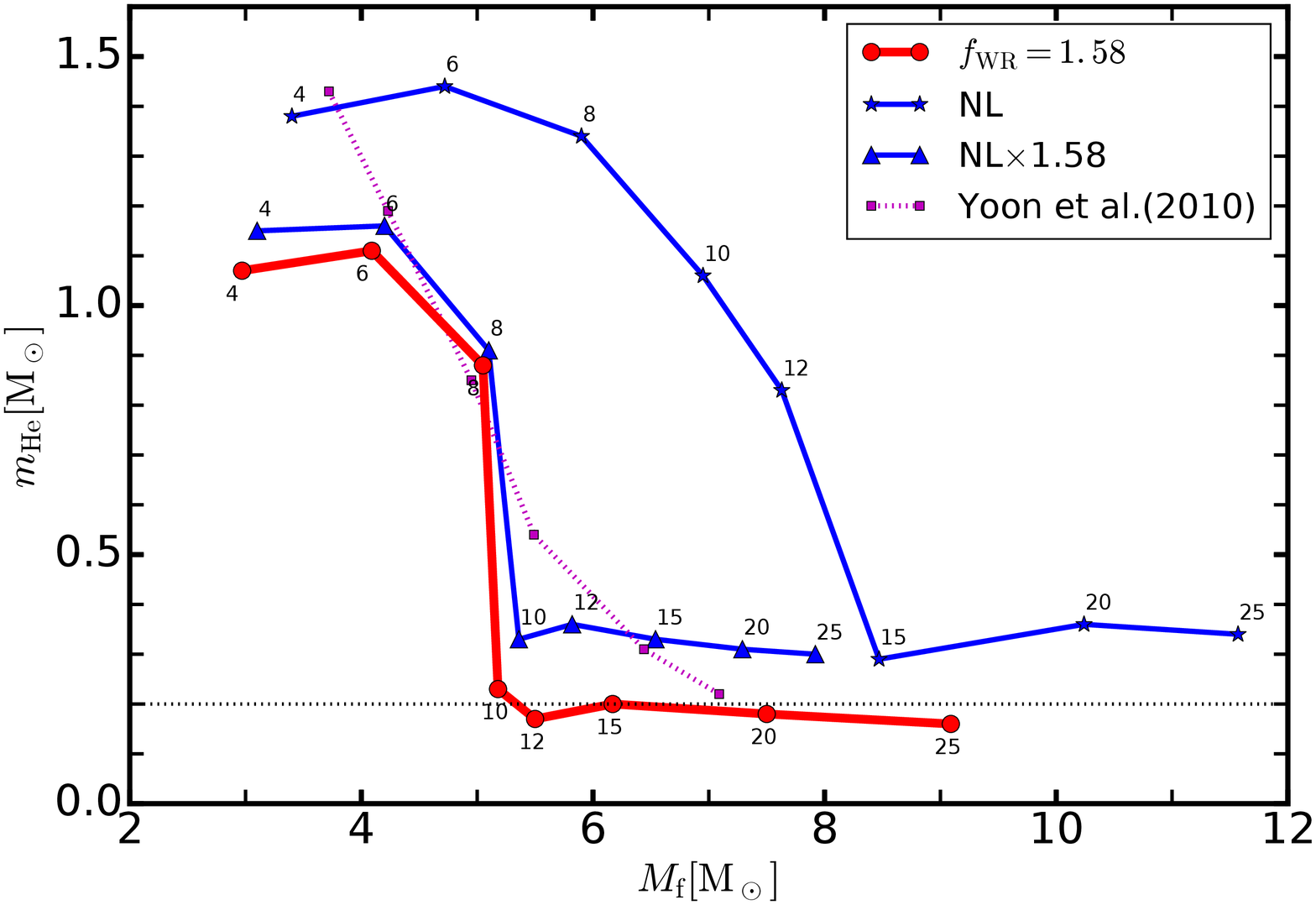}
\includegraphics[width = 0.48\textwidth]{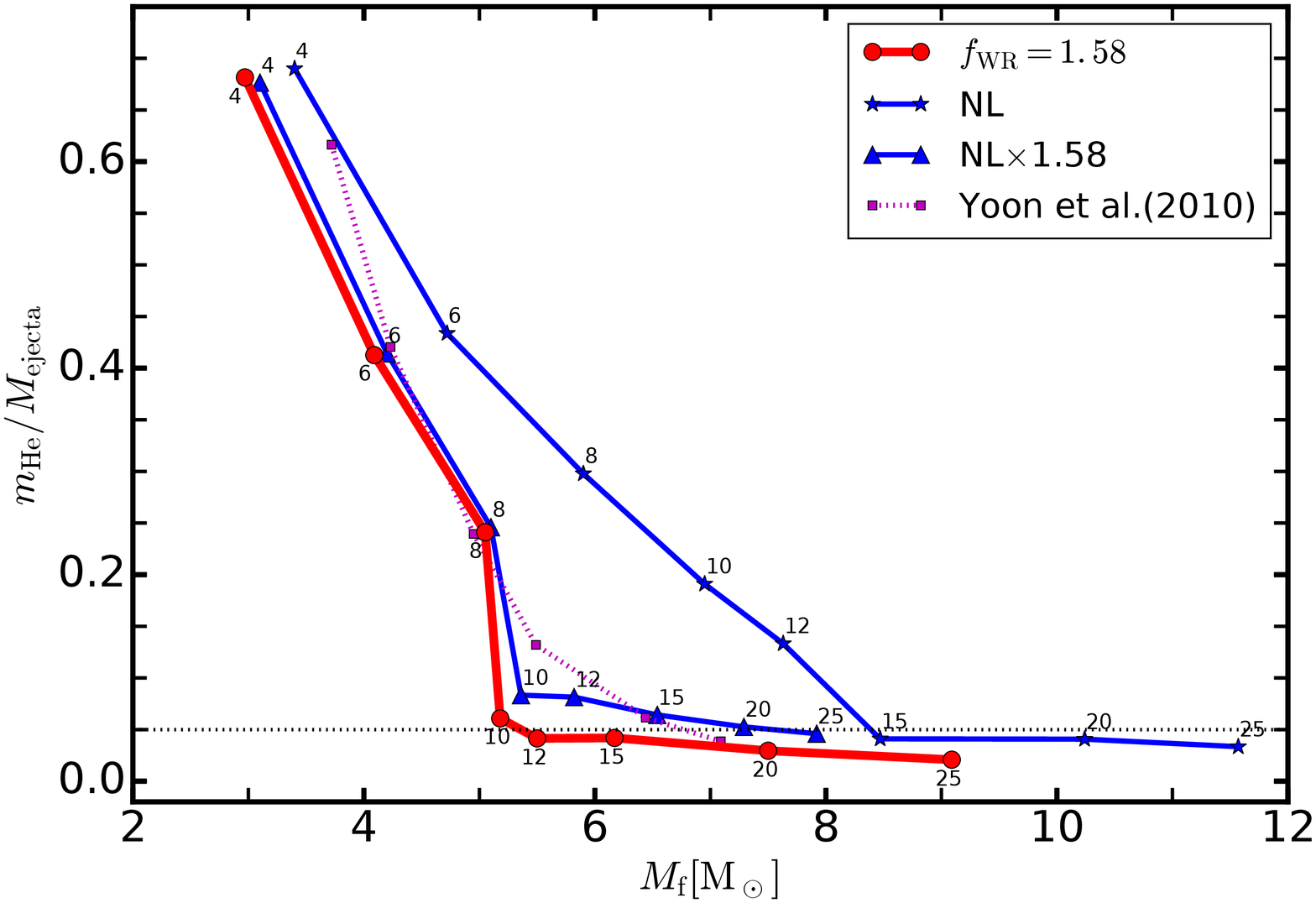}
\includegraphics[width = 0.48\textwidth]{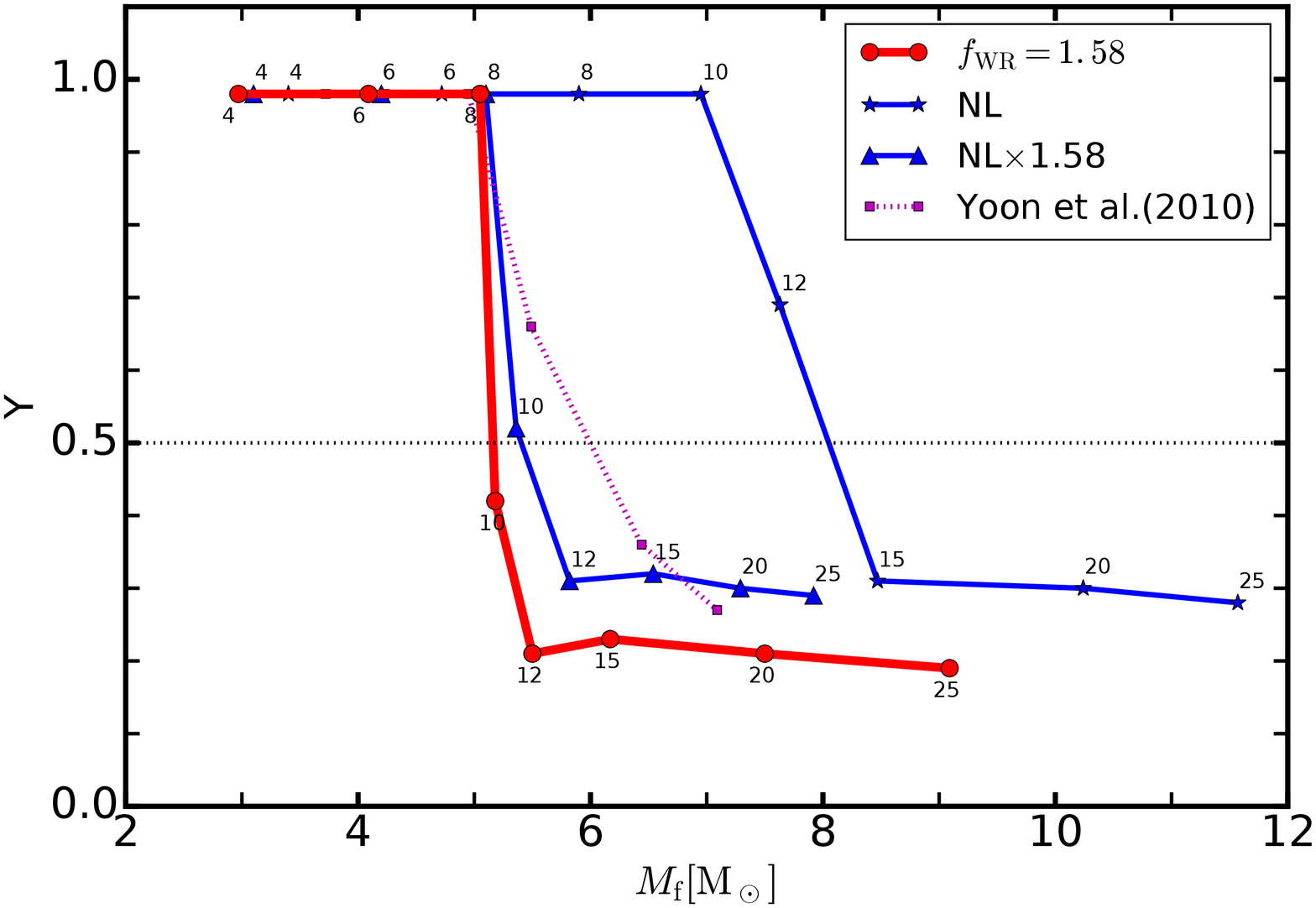}
\caption{\emph{Top panel}: 
Helium mass in the last calculated model as a function of the final mass. 
\emph{Middle panel}: The ratio of helium to ejecta masses as a function of the final mass. 
\emph{Bottom panel}: Surface helium mass fraction in the final model as a function
of the final mass. The red line gives the result with our mass-loss rate prescription 
using $f_\mathrm{WR} = 1.58$ at $Z_\mathrm{init} = Z_\odot$. The blue lines with the star and triangle symbols 
give the results with the NL prescription and $1.58\times$NL prescription at  $Z_\mathrm{init} = Z_\odot$, respectively. 
The number on each symbol denotes the initial mass of the He star. 
The purple dotted line with the filled squares presents the result of He star models at  $Z_\mathrm{init} = Z_\odot$
with $M_\mathrm{He,i} =$ 4, 6, 8, 10, 15, \& 20~$M_\odot$ 
by \citet{Yoon10}, where the WR mass-loss rate prescription by \citet{Hamann95} with 
a reduction factor of 5  was applied.  
\label{fig:snIbc}}
\end{center}
\end{figure}

Our He star models can crudely predict the properties of SN Ib/Ic progenitors
for both single and binary stars as long as $M_\mathrm{He, i} \gtrsim
4.0~M_\odot$ This is because once such a hydrogen-poor star is made either by
stellar wind mass loss from a single star or by mass transfer in a biary
system, the subsequent evolution is governed by WR mass loss. This is
particularly the case for sufficiently high metallicity ($Z\gtrsim Z_\odot$),
for which the WNL phase is relatively short.     Less massive He stars would
undergo Case BB mass transfer in binary systems, in which case full binary star
evolution models are needed to follow the pre-SN
evolution.~\citep[e.g.,][]{Wellstein99, Yoon17}.  

The absence of helium lines in SNe Ic spectra implies that SN Ic progenitors
are helium-poor at the pre-SN stage, while SN Ib progenitors retain  fairly
large amounts of helium. It is still not well known how much helium can be
hidden in SN Ic spectra. Theoretical studies indicate that it depends on the
degree of chemical mixing in the SN ejecta as well as on the progenitor
mass~\citep{Woosley97, Dessart12}.  The upper limit of helium mass in SNe Ic progenitors would
be about 0.14~${M_\odot}$ for well mixed, relatively low-mas SN ejecta
($\sim 1.0~M_\odot$; \citealt{Hashinger12}).  It may reach 
about 1.0~${M_\odot}$ if chemical mixing is only moderate and if the SN
ejecta is sufficiently massive~\citep{Dessart12}, but observations commonly find
signatures of strong mixing in core collapse SN ejecta.  A fairly low helium to ejecta mass ratio
would be therefore a prerequisite for SN Ic progenitors.  A sufficiently low surface
abundance of helium ($Y \lesssim 0.5$) is also needed for the suppression of helium
lines during the early-time phase of a SN Ic~\citep{Dessart11}.

In Fig.~\ref{fig:snIbc}, we present the helium mass in the last calculated models
with $Z_\mathrm{init}={Z_\odot}$. Here the helium mass does not
mean the mass of the helium-rich envelope, but the integrated amount of helium:
$m_\mathrm{He} := \int X_\mathrm{He} dM_r$, where $X_\mathrm{He}$ and $M_r$
denote the local helium mass fraction and the mass coordinate, respectively.  

It is interesting that there exists a very sharp decrease in $m_\mathrm{He}$ at
around $M_\mathrm{f} = 5.0~{M_\odot}$, which corresponds to
$M_\mathrm{He,i} = 8 - 10~{M_\odot}$. This rapid transition results from
the fact that the mass-loss rate increases dramatically when He stars of WN
type ($Y=0.98$) turn into WC stars, which occurs for $M_\mathrm{He,i} \ge 10~{M_\odot}$.

Note also that the predicted helium mass distribution in the case of $f_\mathrm{WR} = 1.58$ 
is clearly bimodal:
$m_\mathrm{He} \approx 1.0~{M_\odot}$ for $M_\mathrm{He,i} < 9.0$ and
$m_\mathrm{He} \approx 0.2~{M_\odot}$ for $M_\mathrm{He,i} > 9.0$.  The
former has $Y = 0.98$ and the latter $Y < 0.5$. The helium to SN ejecta mass
ratio in the latter case is lower than about 0.05.   It is therefore reasonable
to assume that  the former and the latter lead to SNe Ib and Ic, respectively.

The standard NL prescription leads to $M_\mathrm{f} > 8~{M_\odot}$ for SNe Ic
progenitors (assuming $m_\mathrm{He}/M_\mathrm{ejecta} < 0.05$ for SN Ic).
Massive progenitors of $M_\mathrm{f} > 8~{M_\odot}$  would result in too broad
light curves compared to those of ordinary SNe Ic~\citep[e.g.,][]{Dessart17}.
Studies on SN light curves and spectra indicate that ordinary SNe Ic have
ejecta masses less than about 5-6~${M_\odot}$~\citep[e.g.,][]{Drout11, Cano13,
Taddia15, Lyman16}, except for some extreme cases like iPTF15dtg
($M_\mathrm{ejecta} \approx 10~{M_\odot}$; \citealt{Taddia16}).  In this
regard, our result with $f_\mathrm{WR} = 1.58$, which predicts $M_\mathrm{f}
\gtrsim 5~{M_\odot}$ for SN Ic progenitors, is in better agreement with
observations than those with the NL prescription. 

We can see the role of luminosity and $Y$ dependencies of WR mass-loss
rates through the comparison of $f_\mathrm{WR}=1.58$ and 1.58$\times$NL models
in Fig.~\ref{fig:snIbc}.  Both cases  lead to helium poor SN progenitors of
similar final masses for $M_\mathrm{He,i} \ge 10~{M_\odot}$, but the models
with our new mass-loss rate prescription have systematically lower helium masses and
$Y$.  This is because the TSK rate, which we adopt during the WC phase, has
much weaker luminosity and $Y$ dependencies than those of NL.  

The role of the $Y$ dependence is more clearly found when our result is
compared with the He star models by \citet[][Fig.~\ref{fig:snIbc}]{Yoon10}.
They applied the WR mass-loss rate prescription by~\citet{Hamann95} with a
reduction factor of 5. This prescription does not consider the systematic
difference between WNE and WC mass-loss rates,  only having a luminosity dependence of $\dot{M} \propto L^{1.5}$.
Therefore, a He star with this prescription does not undergo mass-loss
enhancement during the WC phase but the mass-loss rate continuously decrease as
the luminosity of the He star decreases due to mass loss.  These models have
final masses similar to those of our new models with $f_\mathrm{WR} = 1.58$.
However,  the helium mass ($m_\mathrm{He}$) in the final model continuously
decreases as a function of the final mass, in contrast to our case where a
bimodal distribution of $m_\mathrm{He}$ is found. 

This comparison illustrates the importance of fine details of the mass-loss
rate prescription for the prediction of SNIb/Ic progenitor properties.  In
particular, we reach the following important conclusion when we consider the
fact that WC stars have systematically stronger mass loss than WNE stars:
\emph{SN Ib and SN Ic  progenitors  are distinctively
different from each other in terms of the helium mass and the surface helium abundance,  
rather than they form a continuous sequence.}

\section{Problem of the temperature discrepancy}\label{sect:temperature}

In our comparison between models and observations on the HR diagram, we
have focused on the luminosity and ignored the temperature discrepancy.  In fact,
our hydrogen-free WR star models predict systematically higher surface
temperatures than the observationally inferred values (Fig.~\ref{fig:hr}
and~\ref{fig:hrlmc}). This discrepancy is found not only with our models, but with
all stellar evolution models~\citep[e.g.,][]{Hamann06, Sander12, Hainich14}.  

Recently, \citet{Graefener12} suggested that the outermost layers of WR stars
may be more inflated  than predicted by standard stellar evolution
models~\citep[e.g.,][]{Ishii99, Petrovic06}, if the opacity in the sub-surface
convective zone is enhanced due to density clumping. \citet{McClelland16} found
that including this clumping effect in WR star models may help reduce the
temperature discrepancy.  They also found that the mass-loss rate of WR stars
should be significantly reduced to have surface temperatures compatible with
observations. This is because the envelope inflation tends to be suppressed
with strong mass loss~\citep{Petrovic06}. 

This leads us to the following dilemma. Increasing the WR mass-loss rate as 
suggested in the present study would alleviate  the luminosity discrepancy (i.e.,
the luminosity range of WC/WO stars) but instead aggravate
the temperature discrepancy, and vice versa.  Compared to our approach that
focuses on the luminosity, \citet{McClelland16} paid more attention to the
surface temperature in their comparison of models and observations. This made
them suggest that hot WO stars are produced by mass loss, and  that relatively
cool WC stars are helium giant stars of relatively low-mass, of which the
surfaces are enriched with carbon by chemical mixing. This
scenario has the following disadvantages. 
\begin{enumerate}
\item Some WC stars are too bright
($\log L/{L_\odot} > 5.5$; Figs~\ref{fig:hr} and~\ref{fig:hrlmc}) to be
explained by helium giants of which the luminosities would be lower than about $\log L/L_\odot = 5.3$~\citep{Yoon12, McClelland16}. 
\item Some WO stars in the LMC are very faint compared to WN
stars ($\log L/{L_\odot} \approx 5.2$; Fig.~\ref{fig:hrlmc}).  We cannot
explain these WO stars by helium giants but need to invoke strong mass loss as
discussed above.  
\item To explain WC stars of $\log L/{L_\odot} \approx
5.0$ with helium giants, we need $M_\mathrm{He,i} \approx 5.0~{M_\odot}$.
He stars of this low mass can be produced only in binary systems. 
But not all of such faint WC stars are found in binary systems~\citep{Sander12}.
\end{enumerate}

More importantly, having an optically thick expanding atmosphere, a WR
star does not have a well defined surface. In the literature, the inner
boundary of the atmosphere model where the optical depth is very large (e.g.,
$\sim 20$ in the Potsdam models) is considered as the surface of the
hydrostatic core of a WR star and the corresponding temperature is compared
with the values given by stellar evolution models.  This means that the
property of this inner boundary  cannot be directly inferred from the spectrum
and certain degeneracy between some model parameters 
is inevitable for very dense winds
~\citep[e.g., see][for detailed discussion]{Hamann06, Groh14}.  On the other
hand, the luminosity measurements are largely limited by the distance
uncertainty.   In the Potsdam Galactic WR sample of Fig.~\ref{fig:hr},
distances to some WR stars are known from their cluster membership.  The
distances to the other WR stars, including the faintest WC stars with $\log
L/L_\odot \lesssim 5.1$,  were inferred from their spectral
types~\citep{Hamann06, Sander12}. This uncertainty should be added as a caveat
to our discussion on Galactic WR stars in Sect.~\ref{sect:evol}.  Future
distance measurements with GAIA would help resolve this issue.  For LMC WR
stars, the distance uncertainty is much smaller and our approach of giving more
weight to the luminosity than to the surface temperature in the comparison
between models and observations may be justified \citep[see also][for a related
discussion]{Sander12}.  We plan to address the issue of envelope inflation and
surface temperature of WR stars in more detail in a forthcoming paper.

\section{Conclusions}\label{sect:conclusion}

We draw the following conclusions from our discussion. 
\begin{itemize}
\item 
The luminosity and initial metallicity dependencies of WNE mass-loss rates are significantly steeper than those of WC/WO mass-loss rates. 
In addition, WC stars have systematically higher mass loss rates than those of WNE stars. 
To properly consider these factors, 
we suggest using a new mass-loss rate prescription that combines the WNE mass-loss rate prescription inferred from the Potsdam WNE sample
and the TSK prescription for WC/WO mass-loss rates. 
\item In addition to these fine details, an overall increase of 
the WR mass-loss rate is needed to explain the formation of relatively 
faint WC/WO stars in our galaxy and LMC (i.e., $\log L/L_\odot < 5.3$). We find that only a moderate 
increase by about 60 per cent (i.e.,  $f_\mathrm{WR} = 1.58$)
is enough to reproduce the observed luminosity range of WC/WO stars, 
which cannot be easily accommodated with the standard NL prescription. 
\item He star models  with this increase of the WR mass-loss rate using our new prescription   can better explain the properties of 
SN Ib/Ic supernova progenitors than those with the standard NL prescription, in terms of final masses and chemical composition. 
We also find a clear bimodal distribution of helium masses and surface helium mass fractions in the new SN progenitor models. 
This implies that the properties of SN Ib and SN Ic progenitors would be distinctively different. 
\end{itemize}

In the present study, we have limited our discussion to the case of
hydrogen-free WR stars. 
In reality, WR stars are born with some amounts of hydrogen left in the
outermost layers, and may spend not a small fraction of the helium burning
lifetime as WNL before they become WNE stars. 
Therefore, our analysis should be extended to WNL stars in
the near future to study  the full evolution from ZAMS to
the pre-SN stage for further
confirmation of our conclusions.  This would also allow us to investigate the
metallicity dependence of the WR population (e.g., WN/WC ratio as a function of
metallicity) and hydrogen-poor SN progenitors, which cannot be properly addressed with
our He star models.

\begin{table*}
\begin{center}
\caption{Physical properties of pure helium star models at solar metallicity ($Z_\mathrm{init}=0.02$)}\label{tab1}
\begin{tabular}{r r r r r r r r r r r r}
\hline
 $M_\mathrm{He,i}$ & $t_\mathrm{evol}$ &  $t_\mathrm{WNE}$ & $t_\mathrm{WC}$  & $M_\mathrm{f}$ & $\log L_\mathrm{f}$ & $\log T_\mathrm{s,f}$ & $M_\mathrm{CO}$ & $m_\mathrm{He}$ & $Y_\mathrm{f}$ & $\log X_\mathrm{C, f}$ & $\log X_\mathrm{O, f}$ \\
\hline
 [$M_\odot$] & [$10^5$ yr]  & [$10^5$ yr]  & [$10^5$ yr]  & [$M_\odot$] &  [$L_\odot$] & [K] & [$M_\odot$] & [$M_\odot$] &  & & \\
\hline
\multicolumn{12}{ c }{$f_\mathrm{WR} = 1.0$}\\
\hline
4.00 &   13.01  &  13.01 & 0.00    &   3.30 &    4.73 &    4.41 &    1.88 &    1.29 &    0.98 &   -3.42 &   -3.47 \\
6.00 &    8.69  &   8.69 & 0.00    &   4.65 &    5.03 &    4.63 &    2.95 &    1.39 &    0.98 &   -3.78 &   -3.16 \\
8.00 &    6.88  &   6.88 & 0.00    &   5.88 &    5.18 &    4.71 &    3.99 &    1.32 &    0.98 &   -3.78 &   -3.16 \\
10.00 &    5.91 &   5.91 &  0.00   &  7.03 &    5.30 &    4.70 &    4.99 &    1.10 &    0.98 &   -3.78 &   -3.16 \\ 
12.00 &    5.32 &   5.19 &  0.13   &    7.99 &    5.37 &    4.90 &    6.02 &    0.70 &    0.83 &   -0.86 &   -2.31 \\
15.00 &    4.78 &   3.48 &  1.29   &    8.80 &    5.41 &    5.13 &    6.83 &    0.25 &    0.32 &   -0.29 &   -0.85 \\
20.00 &    4.23 &   2.15 &  2.08   &   10.96 &    5.55 &    5.31 &    8.65 &    0.27 &    0.25 &   -0.29 &   -0.68 \\
25.00 &    3.88 &   1.54 &  2.34   &   13.28 &    5.67 &    5.30 &   10.50 &    0.26 &    0.23 &   -0.30 &   -0.63 \\
\hline
\multicolumn{12}{ c }{$f_\mathrm{WR} = 1.58$}\\
\hline
 4.00 &   13.76 & 13.76 & 0.00  &    2.97 &    4.68 &    4.28 &    1.79 &    1.07 &    0.98 &   -3.42 &   -3.47 \\
 6.00 &    9.08 &  9.08 & 0.00  &    4.09 &    4.97 &    4.56 &    2.66 &    1.11 &    0.98 &   -3.78 &   -3.16 \\
 8.00 &    7.22 &  7.22 &  0.00 &   5.05  &    5.11 &    4.68 &    3.47 &    0.88 &    0.98 &   -3.78 &   -3.15 \\
10.00 &    6.27 &  5.21 & 1.06  &    5.18 &    5.12 &    4.98 &    3.99 &    0.23 &    0.42 &   -0.33 &   -1.10 \\
12.00 &    5.77 &  3.66 & 2.11  &    5.50 &    5.18 &    5.07 &    4.15 &    0.17 &    0.21 &   -0.27 &   -0.65 \\
15.00 &    5.25 &  2.41 & 2.84  &    6.17 &    5.22 &    5.12 &    4.69 &    0.20 &    0.23 &   -0.27 &   -0.68 \\
20.00 &    4.68 &  1.46 & 3.22  &    7.50 &    5.33 &    5.24 &    5.78 &    0.18 &    0.21 &   -0.28 &   -0.62 \\
25.00 &    4.26 &  1.05 &  3.22 &   9.09  &    5.43 &    5.28 &    7.11 &    0.16 &    0.19 &   -0.29 &   -0.56 \\
\hline
\multicolumn{12}{ c }{NL}\\
\hline
 4.00 &   12.60 & 12.60 & 0.00  &    3.40 &    4.75 &    4.43 &    1.89 &    1.38 &    0.98 &   -3.42 &   -3.47 \\
 6.00 &    8.59 &  8.59 & 0.00  &    4.72 &    5.03 &    4.68 &    3.01 &    1.44 &    0.98 &   -3.78 &   -3.16 \\
 8.00 &    6.83 &  6.38 & 0.00  &    5.90 &    5.16 &    4.75 &    4.01 &    1.34 &    0.98 &   -3.78 &   -3.16 \\
10.00 &    5.91 &  5.91 & 0.00  &    6.95 &    5.30 &    4.71 &    4.99 &    1.06 &    0.98 &   -3.78 &   -3.16 \\
12.00 &    5.32 &  4.92 & 0.40  &    7.63 &    5.35 &    5.02 &    5.85 &    0.53 &    0.69 &   -0.57 &   -1.75 \\
15.00 &    4.86 &  3.12 & 1.68  &    8.47 &    5.42 &    5.08 &    6.46 &    0.29 &    0.31 &   -0.29 &   -0.83 \\
20.00 &    4.33 &  1.85 & 2.48  &   10.24 &    5.52 &    5.29 &    7.98 &    0.36 &    0.30 &   -0.29 &   -0.79 \\
25.00 &    4.05 &  1.27 & 2.78  &   11.57 &    5.58 &    5.28 &    9.11 &    0.34 &    0.28 &   -0.30 &   -0.74  \\
\hline
\multicolumn{12}{ c }{NL$\times1.58$}\\
\hline
4.00 &   13.53 & 13.53  & 0.00  & 3.10 &    4.71 &    4.35 &    1.85 &    1.15 &    0.98 &   -3.42 &   -3.47 \\
6.00 &    8.99 &  8.99  & 0.00  & 4.20 &    4.97 &    4.60 &    2.72 &    1.16 &    0.98 &   -3.78 &   -3.16 \\
8.00 &    7.21 &  7.21 &  0.00  & 5.10 &    5.12 &    4.68 &    3.49 &    0.91 &    0.98 &   -3.78 &   -3.16 \\
10.00 &    6.28 & 5.15  & 1.13 &  5.36 &    5.14 &    5.03 &    4.04 &    0.33 &    0.52 &   -0.39 &   -1.31 \\
12.00 &    5.80 & 3.49  & 2.31 &  5.82 &    5.21 &    4.98 &    4.24 &    0.26 &    0.31 &   -0.28 &   -0.87 \\
15.00 &    5.32 & 2.19  & 3.12 &  6.54 &    5.28 &    5.00 &    4.84 &    0.33 &    0.32 &   -0.29 &   -0.86 \\
20.00 &    4.88 & 1.26  & 3.62 &  7.29 &    5.34 &    5.07 &    5.47 &    0.31 &    0.30 &   -0.29 &   -0.81 \\
25.00 &    4.64 & 0.88  & 3.77 &  7.92 &    5.38 &    5.11 &    6.03 &    0.30 &    0.29 &   -0.29 &   -0.76 \\
\hline
\end{tabular}
\end{center}
{
Each column has the following meaning. $M_\mathrm{He,i}$: the initial mass of the helium star, $t_\mathrm{evol}$: the whole evolutionary time
of the helium star, $t_\mathrm{WNE}$: the lifetime of the WNE phase, $t_\mathrm{WC}$: the lifetime of the WC phase, $M_\mathrm{f}$: the total mass 
of the last calculated model (i.e., the final mass), $L_\mathrm{f}$ : the luminosity of the last calculated model,  $T_\mathrm{s,f}$: the surface
temperature of the last calculated model, $M_\mathrm{CO}$: the CO core mass in the last calculated model, $m_\mathrm{He}$: the integrated helium 
mass in the last calculated model (i.e., $m_\mathrm{He} = \int X_\mathrm{He} dM_r$), $Y_\mathrm{f}$: the surface mass fraction of helium 
in the last calculated model,  $X_\mathrm{C,f}$: the surface mass fraction of carbon 
in the last calculated model,  $X_\mathrm{O,f}$: the surface mass fraction of oxygen in the last calculated model.  
}
\end{table*}

\begin{table*}
\begin{center}
\caption{Physical properties of pure helium star models at LMC metallicity ($Z_\mathrm{init}=0.01$)}\label{tab2}
\begin{tabular}{r r r r r r r r r r r r}
\hline
 $M_\mathrm{He,i}$ & $t_\mathrm{evol}$ &  $t_\mathrm{WNE}$ & $t_\mathrm{WC}$  & $M_\mathrm{f}$ & $\log L_\mathrm{f}$ & $\log T_\mathrm{s,f}$ & $M_\mathrm{CO}$ & $m_\mathrm{He}$ & $Y_\mathrm{f}$ & $\log X_\mathrm{C, f}$ & $\log X_\mathrm{O, f}$ \\
\hline
 [$M_\odot$] & [$10^5$ yr]  & [$10^5$ yr]  & [$10^5$ yr]  & [$M_\odot$] &  [$L_\odot$] & [K] & [$M_\odot$] & [$M_\odot$] &  & & \\
\hline
\multicolumn{12}{ c }{$f_\mathrm{WR} = 1.58$}\\
\hline
10.00 &    6.06 &  6.06 & 0.00 &   6.90 &    5.29 &    4.98 &    4.93 &    1.03 &    0.99 &   -4.08 &   -3.46 \\
12.00 &    5.44 &  5.08& 0.35 &   7.51 &    5.32 &    5.23 &    5.85 &    0.45 &    0.63 &   -0.48 &   -1.55 \\
15.00 &    4.94 &  3.41 & 1.53 &   8.08 &    5.37 &    5.27 &    6.26 &    0.17 &    0.19 &   -0.28 &   -0.57 \\
20.00 &    4.43 &  2.09 & 2.34 &   9.64 &    5.47 &    5.31 &    7.54 &    0.19 &    0.21 &   -0.28 &   -0.59 \\
25.00 &    4.10 &  1.51 & 2.59 &  11.39 &    5.57 &    5.31 &    9.01 &    0.20 &    0.19 &   -0.30 &   -0.54 \\
\hline
\multicolumn{12}{ c }{NL}\\
\hline
10.00 &    5.92 & 5.92 & 0.00 &   7.58 &    5.33 &    5.00 &    5.42 &    1.37 &    0.99 &   -4.08 &   -3.47 \\
12.00 &    5.32 & 5.32 & 0.00 &   8.73 &    5.44 &    5.01 &    6.51 &    1.11 &    0.99 &   -4.08 &   -3.47 \\
15.00 &    4.77 & 4.12 & 0.65 &   9.52 &    5.44 &    5.27 &    7.47 &    0.26 &    0.41 &   -0.32 &   -1.01 \\
20.00 &    4.30 & 2.46 & 1.84 &  11.49 &    5.56 &    5.26 &    9.05 &    0.35 &    0.25 &   -0.29 &   -0.67 \\
25.00 &    4.02 & 1.69 & 2.33 &  12.79 &    5.64 &    5.29 &   10.22 &    0.33 &    0.27 &   -0.29 &   -0.68 \\
\hline
\end{tabular}
\end{center}
\end{table*}

\section*{Acknowledgements}
The author is grateful to Andreas Sander, who refereed this paper, for
his constructive comments,  to Frank Tramper for helpful discussion, to JJ Eldridge
for his careful reading of the manuscript and  useful comments, and to
Alexander Heger and the Monash Center for Astrophysics for the support by 
the distinguished vistor program.  This work was supported by the Korea Astronomy
and Space Science Institute under the R\&D program (Project No. 3348- 20160002)
supervised by the Ministry of Science, ICT and Future Planning.

\end{document}